\documentclass{prop2015mod}
\usepackage[english]{babel}

\setcopyrightyear{2016}%
\DOIprefix{10.1002}%
\DOIsuffix{prop.201400xxx}%
\Volume{}%
\Issue{}%
\Month{May}%
\Year{2016}%
\Receiveddate{May 2, 2016}
\title{Current Noise in Tunnel Junctions}
\author[Frey]{Moritz Frey\inst{1,}}
\author[Grabert]{Hermann Grabert\inst{1}\footnote{Corresponding author\quad E-mail:~\textsf{grabert@uni-freiburg.de}}}
\address[1]{Physikalisches Institut, Universit\"at
Freiburg, Hermann-Herder-Stra{\ss}e 3, 79104 Freiburg, Germany}
\begin{abstract}
We study current fluctuations in tunnel junctions driven by a voltage source. The voltage is applied to the tunneling element via an impedance providing an electromagnetic environment of the junction. We use circuit theory to relate the fluctuations of the current flowing in the leads of the junction with the voltage fluctuations generated by the environmental impedance and the fluctuations of the tunneling current. The spectrum of current fluctuations is found to consist of three parts: a term arising from the environmental Johnson-Nyquist noise, a term due to the shot noise of the tunneling current and a third term describing the cross-correlation between these two noise sources. Our phenomenological theory reproduces previous results based on the Hamiltonian model for the dynamical Coulomb blockade  and provides a simple understanding of the current fluctuation  spectrum in terms of circuit theory and properties of the average current. Specific results are given for a tunnel junction driven through a resonator.
\end{abstract}
\shortabstract
\begin{document}
\renewcommand{\today}[1]{{\bf XX},xxx (2016)}
\maketitle

\section{Introduction}\label{sec:one}
Recent years have seen a revival of experimental studies of tunnel junctions coupled to an electromagnetic environment\cite{Hofheinz_2011,Altimiras_2014,Parlavecchio_2015}. In these experiments the electromagnetic environment is designed to display a pronounced resonance mode and the devices are partly driven by voltages at microwave frequencies. Along with these new experimental studies, the theory of the dynamical Coulomb blockade (DCB) developed in the 1990ties\cite{Devoret_1990,Girvin_1990,Grabert_1991,Ingold_1992,Lee_1996} to describe the influence of the electromagnetic environment on tunneling was reconsidered and extended to ac driven devices\cite{Safi_2011,Souquet_2013,Grabert_2015,Roussel_2016,Mora_2016,Frey_2016}.

In this paper we combine circuit theory with results of the DBC theory for the average current to examine the spectrum of current fluctuations in the leads of the circuit. As pointed out by Landauer and Martin\cite{Landauer_1991}, the tunneling current studied in most of the theoretical papers is not directly observable. Measurable quantities are typically related to the current flowing in the leads of the junction. In Sect.~\ref{sec:two} we present the model and use circuit theory to express the fluctuations of the current drawn from the voltage source in terms of the fluctuations of the tunneling current and the Johnson-Nyquist voltage fluctuations produced by the environmental impedance.

In Sect.~\ref{sec:three} the spectral function of current fluctuations is introduced and evaluated. We find that the spectrum consists of three parts associated with the Johnson-Nyquist noise of the environmental impedance, the shot noise of the tunneling element and a third contribution related to the cross-correlation between these two noise sources. While the circuit theoretical considerations are rather generally valid for tunneling elements, explicit expressions for all three noise contributions are obtained by making use of results of the DCB theory for a tunnel junction in the weak tunneling limit. In Sect.~\ref{sec:four} the theory is illustrated by applying it to a tunnel junction driven through a resonator. Finally, Sect.\ref{sec:five} contains our conclusions.

\section{Current fluctuations and circuit theory}\label{sec:two}
\begin{figure}
\centering
\includegraphics[width=0.4\textwidth]{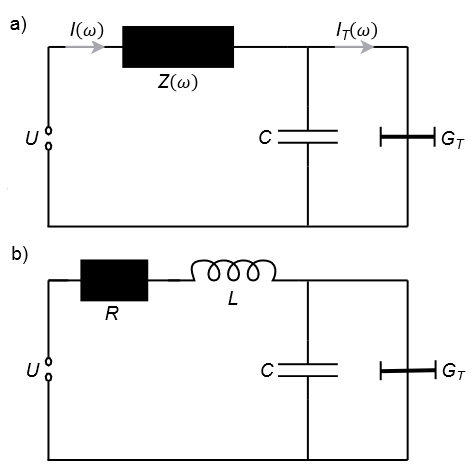}
\caption{\label{fig1} a) Circuit diagram of a voltage biased tunneling element.  The setup shows a tunnel junction with capacitance $C$ and tunneling conductance $G_T$ coupled to an external voltage source $U$ via a lead impedance $Z(\omega)$. The current $I$ flowing through the environmental impedance and the tunneling current $I_T$ are also indicated. b) Circuit diagram of a tunnel junction driven through an $LC$ resonator with lead resistance $R$.}
\end{figure}
We consider the circuit depicted in Fig.~\ref{fig1}a showing a tunneling element with tunneling conductance $G_T$ in series with an impedance $Z$ providing an electromagnetic environment of the junction. An external voltage $U$ is applied to the device. The current drawn from the voltage source is denoted by $I$ and the tunneling current across the junction by $I_T$.  The junction capacitance $C$ is charged [discharged] by the current $I$ [$I_T$]  so that the time rate of change of the junction charge $Q$ obeys
\begin{equation}\label{dotQ}
\dot Q= I-I_T
\end{equation}
Hence, at constant applied voltage $U$ the average currents coincide, i.e., 
$
\langle I \rangle  = \langle I_T  \rangle
$,
and they are given by the result of the DCB theory \cite{Devoret_1990,Girvin_1990,Grabert_1991,Ingold_1992}. This average current will be denoted by $I(U)$ in the sequel.

Here, we study the fluctuations $\delta I=I-\langle I \rangle$ of the current $I$ and of other circuit variables in Fourier space.  Eq.~(\ref{dotQ}) then reads 
\begin{equation}\label{omQ1}
-i\omega \,\delta Q(\omega)=\delta I(\omega) - \delta I_T(\omega)
\end{equation}
The voltage $V$ across the environmental impedance may be written as
\begin{equation}\label{Vom}
V(\omega)=Z(\omega) I(\omega) + \delta v_N(\omega)
\end{equation}
where $\delta v_N$ is the noise voltage generated by the electromagnetic environment. This  Johnson-Nyquist noise has the properties\cite{Johnson_1932,Nyquist_1932,Clerk_2010}
\begin{equation}\label{vNav}
\langle \delta v_N(\omega) \rangle =0
\end{equation}
and
\begin{equation}\label{vNvar}
\langle \delta v_N(\omega)\delta v_N(-\omega) \rangle =\frac{2\hbar\omega}{1-e^{-\beta\hbar\omega}} Z^{\prime}(\omega)
\end{equation}
where $\beta=1/k_BT$ is the inverse temperature and $Z^{\prime}$ the real part of $Z$.

Since the charge $Q$ is related to the voltage $V_J$ across the junction by $V_J=Q/C$, we have
\begin{equation}
\delta V_J(\omega)=\frac{\delta Q(\omega)}{C}
\end{equation}
On the other hand, the voltage fluctuations obey
\begin{equation}\label{sumV}
\delta V(\omega) + \delta V_J(\omega) =0
\end{equation}
which implies
\begin{equation}\label{omQ2}
-i\omega\, \delta Q(\omega)= -i\omega C\, \delta V_J(\omega) = i\omega C\, \delta V(\omega)
\end{equation}
From Eqs.~(\ref{Vom}) and (\ref{vNav}) we obtain for the voltage fluctuations across the environmental impedance
\begin{equation}
\delta V(\omega) = Z(\omega)\delta I(\omega)+\delta v_N(\omega)
\end{equation}
which may be inserted into Eq.~(\ref{omQ2}) to yield
\begin{equation}
-i\omega\, \delta Q(\omega)=i\omega C[Z(\omega)\delta I(\omega)+\delta v_N(\omega)]
\end{equation}
When this is combined with Eq.~(\ref{omQ1}), we can eliminate the charge fluctuations $\delta Q$ and obtain
\begin{equation}\label{deltaI}
[1-i\omega C Z(\omega)]\delta I(\omega)= \delta I_T(\omega) +i\omega C  \delta v_N(\omega)
\end{equation}
Introducing the total impedance of the electromagnetic environment\cite{Ingold_1992}
\begin{equation}\label{Zt}
Z_t(\omega)=\frac{1}{Y(\omega)-i\omega C}
\end{equation}
where
$
Y(\omega)=1/Z(\omega)
$
is the admittance,
Eq.~(\ref{deltaI}) may be transformed to read
\begin{equation}\label{IN+T}
\delta I(\omega)= Y(\omega) Z_t(\omega)[\delta I_T(\omega) +i\omega C  \delta v_N(\omega)]
\end{equation}
which expresses the current fluctuations observed in the leads of the junction in terms of the noise voltage $\delta v_N$ of the electromagnetic environment and the fluctuations $\delta I_T$ of the tunneling current.

\section{Current noise}\label{sec:three}

The spectral function of current fluctuations is defined by
\begin{equation}
S(\omega)=\langle \delta I(\omega)\delta I(-\omega)\rangle
\end{equation}
Inserting the representation (\ref{IN+T}) of $\delta I(\omega)$, we obtain 
\begin{eqnarray}
&&S(\omega) =\left\vert Y(\omega) Z_t(\omega)\right\vert^2 \big[\omega^2 C^2 \langle \delta v_N(\omega)\delta v_N(-\omega)\rangle\\ \nonumber
&&\qquad+\langle\delta I_T(\omega) \delta I_T(-\omega)\rangle \\ \nonumber
&&\qquad +i\omega C \big( \langle \delta v_N(\omega) \delta I_T(-\omega)\rangle
- \langle \delta I_T(\omega)\delta v_N(-\omega)\rangle  \big)\big]
\end{eqnarray}
where we have made use of the symmetries $Y(-\omega)=Y(\omega)^*$ and $Z_t(-\omega)=Z_t(\omega)^*$.
Hence, the spectral function is a sum of three contributions\cite{Lee_1996,Frey_2016}
\begin{equation}
S(\omega)=S_N(\omega)+S_T(\omega)+S_{NT}(\omega)
\end{equation}
where
\begin{equation}\label{SN}
S_N(\omega)= \left\vert \omega C Y(\omega) Z_t(\omega)\right\vert^2 \langle \delta v_N(\omega)\delta v_N(-\omega)\rangle
\end{equation}
\begin{equation}\label{ST}
S_T(\omega)=\left\vert Y(\omega)Z_t(\omega)\right\vert^2 \langle\delta I_T(\omega) \delta I_T(-\omega)\rangle 
\end{equation}
and
\begin{equation}\label{SNT}
S_{NT}(\omega)=\left\vert Y(\omega) Z_t(\omega)\right\vert^2 \omega C 
\big(-i  \langle   \delta I_T(\omega) \delta v_N(-\omega)\rangle + \hbox{c.c.}\big)
\end{equation}
We now discuss these contributions separately.

\subsection{Johnson-Nyquist noise}

The contribution (\ref{SN}) to the spectral function is a consequence of the environmental Johnson-Nyquist noise (\ref{vNvar}). 
Combining Eqs.~(\ref{vNvar}) and (\ref{SN}) we obtain
\begin{equation}\label{SN2}
S_N(\omega)= \big\vert \omega C Z_t(\omega)\big\vert^2 \frac{2\hbar\omega}{1-e^{-\beta\hbar\omega}} Y^{\prime}(\omega)
\end{equation}
where we have made use of the relation
\begin{equation}
\vert Y(\omega)\vert^2 Z^{\prime}(\omega)=Y^{\prime}(\omega)
\end{equation}
The result (\ref{SN2}) coincides with findings\cite{Frey_2016} based on a Hamiltonian model. Note that the contribution $S_N(\omega)$ to the current noise in the outer circuit differs from the standard Johnson-Nyquist current noise $2\hbar\omega Y^{\prime} (\omega)/[1-e^{-\beta\hbar\omega}]$ of an admittance $Y(\omega)$ \cite{Johnson_1932,Nyquist_1932,Clerk_2010} by the absolute value squared of a transmission factor 
\begin{equation}\label{HC}
H_C(\omega)=\frac{-i\omega C}{Y(\omega)-i\omega C}=-i\omega C Z_t(\omega)
\end{equation}
This factor results from the fact that the noise generated at the environmental impedance has to be transmitted to the outer circuit via the junction capacitance $C$.

\subsection{Shot noise}

The second part $S_T(\omega)$ of the noise spectrum in the outer circuit is related to the shot noise of the tunneling current
\begin{equation}
S_{TT}(\omega)=\langle\delta I_T(\omega) \delta I_T(-\omega)\rangle
\end{equation}
The contribution (\ref{ST}) to the spectral function may be written as
\begin{equation}\label{ST2}
S_T(\omega) =\big\vert Y(\omega)Z_t(\omega)\big\vert^2 S_{TT}(\omega)
\end{equation}
Again, the observable noise in the leads differs from the shot noise at the tunnel junction by the absolute value squared of a transmission factor
\begin{equation}\label{HZ}
H_Z(\omega)=\frac{Y(\omega)}{Y(\omega)-i\omega C}=Y(\omega) Z_t(\omega)=1-H_C(\omega)
\end{equation}
arising from the fact that the noise has to be transmitted to the outer circuit via the environmental impedance $Z(\omega)$.

The result (\ref{ST2}) can be combined with the expression for the shot noise of the tunneling current of a junction in the DCB regime\cite{Altimiras_2014}
\begin{equation}\label{STT}
S_{TT}(\omega)=e\left[ \frac{I(U+\hbar\omega/e)}{1-e^{-\beta(eU+\hbar\omega)}}
+\frac{I(U-\hbar\omega/e)}{e^{\beta(eU-\hbar\omega)}-1}\right]
\end{equation}
where
$
I(U)=\langle I \rangle =\langle I_T\rangle
$
is the average current in the presence of an applied voltage $U$. Combining Eqs.~(\ref{ST2}) and (\ref{STT}), we obtain for the component $S_T(\omega)$ of the spectral function likewise an expression which is in  accordance with calculations based on the Hamiltonian model of the DCB theory\cite{Frey_2016}.

\subsection{Cross-correlation noise}

The third component (\ref{SNT}) of the spectral function involves the correlation
$\langle \delta I_T(\omega)  \delta v_N(-\omega) \rangle $ between voltage fluctuations $\delta v_N$ generated by the electromagnetic environment and fluctuations $\delta I_T$ of the tunneling current.
To analyze this contribution we first note that in view of Eq.~(\ref{sumV})	the fluctuations $\delta V_J$ of the voltage across the tunnel junction may be expressed in the form
\begin{eqnarray}\label{delVJ}
\delta V_J(\omega) &=& - \delta V(\omega) = - Z(\omega) \delta I(\omega) -\delta v_N(\omega)
 \\ \nonumber
&=& - Z_t(\omega) [\delta I_T(\omega) +Y(\omega) \delta v_N(\omega)]
\end{eqnarray}
where we have employed Eqs.~(\ref{Vom}) and (\ref{IN+T}) and made use of $1+i\omega CZ_t(\omega)=Y(\omega)Z_t(\omega)$ to get the last expression.

We further note that the response of the average tunneling current in the presence of an applied dc voltage $U$ to a small alternating voltage $\delta V_J(\omega)$ of frequency $\omega$ can be written as
\begin{equation}\label{ITf}
\delta \langle I_T \rangle =  Y_J(U,\omega)\, \delta V_J(\omega)
\end{equation}
where
$
Y_J(U,\omega)
$
is the admittance of the junction. This quantity can be determined from the expression for the average tunneling current in the presence of dc and ac voltages\cite{Parlavecchio_2015,Grabert_2015}. 
As shown in the Appendix, one has
\begin{eqnarray}\label{YJ}
&&Y_J(U,\omega) =\frac{e}{2\hbar\omega}\big\{   I(U+\hbar\omega/e) - I(U-\hbar\omega/e)
 \\ \nonumber
&&\qquad -i  \left[ I_{KK}(U+\hbar\omega/e)
+I_{KK}(U-\hbar\omega/e) -2I_{KK}(U) \right]\big\}
\end{eqnarray}
where $I_{KK}(U)$ is the Kramers-Kronig transform of the current voltage characteristic $I(U)$.

We now want to determine the fluctuation $\delta I_T(\omega)$ of the tunneling current caused by a Johnson-Nyquist voltage fluctuation $\delta v_N(\omega)$. According to the chain rule we may write
\begin{equation}
\frac{\partial I_T(\omega)}{\partial v_N(\omega)} = \frac{\partial I_T(\omega)}{\partial V_J(\omega)}
\frac{\partial V_J(\omega)}{\partial v_N(\omega)}\Bigg\vert_{I_T}
\end{equation}
where the first factor on the rhs is the admittance $Y_J(U,\omega)$ of the junction while Eq.~(\ref{delVJ}) gives for the second factor
\begin{equation}
\frac{\delta V_J(\omega)}{\delta v_N(\omega)}\Bigg\vert_{I_T}=-Y(\omega) Z_t(\omega)
\end{equation}
Hence, we have
\begin{eqnarray}\label{ITvN}
&&\langle \delta I_T(\omega)\delta v_N(-\omega)\rangle \\ \nonumber
&&\quad = \Bigg\langle \frac{\partial I_T(\omega)}{\delta v_N(\omega)}
 \delta v_N(\omega) \delta v_N(-\omega) \Bigg\rangle \\ \nonumber
&&\quad = - Y_J(U,\omega)\,Y(\omega)Z_t(\omega)  \langle  \delta v_N(\omega) \delta v_N(-\omega) \rangle
\end{eqnarray}
Combining Eqs.~(\ref{vNvar}), (\ref{YJ}) and (\ref{ITvN}), we obtain for the cross-correlation noise (\ref{SNT})
\begin{eqnarray}\label{SNT2}
&&S_{NT}(\omega)= \frac{e}{1-e^{-\beta\hbar\omega}}\left\vert Y(\omega) Z_t(\omega)\right\vert^2 Z^{\prime}(\omega)\,\omega C \\ \nonumber
&&\quad\times \Big\{i  Y(\omega)Z_t(\omega)\big[   I(U+\hbar\omega/e) - I(U-\hbar\omega/e)  \\ \nonumber
&&\qquad -i  \big( I_{KK}(U+\hbar\omega/e)
+I_{KK}(U-\hbar\omega/e) -2I_{KK}(U) \big)\big] \\ \nonumber
&&\qquad + \hbox{c.c.}\, \Big\}
\end{eqnarray}

It is now convenient to introduce the polar decomposition\cite{Grabert_2015,Frey_2016}
\begin{equation}\label{polar}
H_Z(\omega) = Y(\omega)Z_t(\omega) =\Xi(\omega)\, e^{i\eta(\omega)}
\end{equation}
of the transmission factor (\ref{HZ}) into modulus $\Xi(\omega)$ and phase $\eta(\omega)$. The imaginary part of the transmission factor then reads
\begin{eqnarray}
\Xi(\omega)\sin[\eta(\omega)]&=&\omega C Y^{\prime}(\omega) \vert Z_t(\omega)\vert^2 
\\ \nonumber
&=&\omega C Z^{\prime}(\omega) \vert Y(\omega)Z_t(\omega)\vert^2
\end{eqnarray}
In terms of $\Xi(\omega)$ and $\eta(\omega)$, the result (\ref{SNT2}) takes the form
\begin{eqnarray}\label{SNT3}
&&S_{NT}(\omega)= \frac{e}{1-e^{-\beta\hbar\omega}}\Xi(\omega)\sin[\eta(\omega)]\\ \nonumber
&&\quad\times \Big\{i  \Xi(\omega)\left(\cos[\eta(\omega)]+i\sin[\eta(\omega)]\right)
\\ \nonumber
&&\qquad\times\big[   I(U+\hbar\omega/e) - I(U-\hbar\omega/e)  \\ \nonumber
&&\qquad -i  \big( I_{KK}(U+\hbar\omega/e)
+I_{KK}(U-\hbar\omega/e) -2I_{KK}(U) \big)\big] \\ \nonumber
&&\qquad + \hbox{c.c.}\, \Big\}
\end{eqnarray}
which gives
\begin{eqnarray}\label{SNT4}
&&S_{NT}(\omega)=\frac{2e}{1-e^{-\beta\hbar\omega}}\Xi(\omega)^2 \\ \nonumber
&&\quad\times\Big\{ \sin^2[\eta(\omega)]\big[   I(U-\hbar\omega/e) - I(U+\hbar\omega/e) \big] \\ \nonumber
&&\qquad + \frac{\sin[2\eta(\omega)]}{2} \big[ I_{KK}(U+\hbar\omega/e)  \\ \nonumber
&&\qquad
+I_{KK}(U-\hbar\omega/e) -2I_{KK}(U) \big]  \Big\}
\end{eqnarray}
This expression for the cross-correlation noise is again in accordance with the result\cite{Frey_2016} of a calculation based on the Hamiltonian model of the DCB theory.

\section{Tunnel junction driven through a resonator}\label{sec:four}
We now examine a tunnel junction driven through an $LC$-resonator. This problem has been studied recently\cite{Altimiras_2014,Parlavecchio_2015,Grabert_2015,Mora_2016,Frey_2016} both experimentally and theoretically. 

\subsection{Model parameters}

The circuit diagram depicted in Fig.~\ref{fig1}b shows an environmental impedance of the form
\begin{equation}\label{ZLC}
Z(\omega)=R-i\omega L
\end{equation}
with an Ohmic lead resistance $R$ and an inductance $L$. The resonance frequency of the $LC$-resonator is 
$
\omega_0=1/\sqrt{LC}
$
and it has the characteristic impedance 
$
Z_c=\sqrt{L/C}
$
implying a quality factor $Q_f=Z_c/R$. We shall also use the loss factor
$
\gamma=1/Q_f
$.

For this circuit the transmission factors (\ref{HC}) and (\ref{HZ}) read
\begin{equation}\label{HCLC}
H_C(\omega)
=-\frac{\omega^2 +i\gamma\omega_0\omega }{\omega_0^2-\omega^2-i\gamma\omega_0\omega }
\end{equation}
and
\begin{equation}\label{HZLC}
H_Z(\omega)=\frac{\omega_0^2}{\omega_0^2-\omega^2-i\gamma\omega_0\omega}
\end{equation}
which implies a modulus
\begin{equation}\label{XiLC}
\Xi(\omega)=\frac{\omega_0^2}{\sqrt{\left(\omega_0^2-\omega^2\right)^2+\left(\gamma\omega_0\omega\right)^2}} 
\end{equation}
and a phase
\begin{equation} \label{etaLC}
\eta(\omega)=\arctan\left(\frac{\gamma\omega_0\omega}{\omega_0^2-\omega^2}\right)
\end{equation}
where the values of $\arctan$ are to be chosen in the interval $[0,\pi)$. 
We then have
\begin{equation}\label{sineta}
\sin[\eta(\omega)]=\frac{\gamma\omega_0\omega}{\sqrt{(\omega_0^2-\omega^2)^2+(\gamma\omega_0\omega)^2}}
\end{equation}
and
\begin{equation}\label{sin2eta}
\sin[2\eta(\omega)]= \frac{2\gamma\omega_0\omega(\omega_0^2-\omega^2)}{(\omega_0^2-\omega^2)^2+(\gamma\omega_0\omega)^2}
\end{equation}

\subsection{Spectral function}

From Eq.~(\ref{ZLC}) we obtain for the real part of the environmental admittance
\begin{equation}\label{YprimeLC}
Y^{\prime}(\omega)=\frac{\gamma^2\omega_0^2}{\gamma^2\omega_0^2+ \omega^2}\frac{1}{R}
\end{equation}
The expressions (\ref{HCLC}) and (\ref{YprimeLC}) may now be inserted into Eq.~(\ref{SN2}) to yield for the Johnson-Nyquist part of the spectral function
\begin{equation}\label{SNLC}
S_N(\omega)= \frac{2\hbar\omega}{1-e^{-\beta\hbar\omega}}
\frac{\gamma^2\omega_0^2 \omega^2(\omega^2 +\gamma^2\omega_0^2)}{\left[(\omega_0^2-\omega^2)^2+\gamma^2\omega_0^2\omega^2\right]  \left(\omega^2+\gamma^2\omega_0^2\right)}\frac{1}{R}
\end{equation}
The shot noise part (\ref{ST2}) of the spectrum reads
\begin{eqnarray}\label{STLC}
S_T(\omega) &=& \frac{e\omega_0^4}{\left(\omega_0^2-\omega^2\right)^2+\left(\gamma\omega_0\omega\right)^2} \\ \nonumber
&&\quad\times \left[ \frac{I(U+\hbar\omega/e)}{1-e^{-\beta(eU+\hbar\omega)}}
+\frac{I(U-\hbar\omega/e)}{e^{\beta(eU-\hbar\omega)}-1}\right]
\end{eqnarray}
where we have made use of Eqs.~(\ref{STT}) and (\ref{XiLC}).
Finally, for the cross-correlation part (\ref{SNT4}) of the spectrum we find by virtue of Eqs.~(\ref{XiLC}), (\ref{sineta}) and (\ref{sin2eta})
\begin{eqnarray}\label{SNTLC}
&&S_{NT}(\omega)=\frac{e}{1-e^{-\beta\hbar\omega}}\frac{2\gamma\omega_0^5\omega}{\left[\left(\omega_0^2-\omega^2\right)^2+\left(\gamma\omega_0\omega\right)^2\right]^2} \\ \nonumber
&&\quad\times\Big\{
\gamma\omega_0\omega\big[   I(U-\hbar\omega/e) - I(U+\hbar\omega/e) \big] \\ \nonumber
&&\qquad + (\omega_0^2-\omega^2) \big[ I_{KK}(U+\hbar\omega/e) +I_{KK}(U-\hbar\omega/e)  \\ \nonumber
&&\qquad
 -2I_{KK}(U) \big]  \Big\}
\end{eqnarray}

\subsection{Results for specific parameters}
To illustrate these results, we now consider an $LC$-cicuit with quality factor $Q_f=5$. A general method to calculate the current voltage characteristic $I(U)$ for a tunnel junction in an electrodynamic environment  has been presented previously \cite{Ingold_1991}. The approach is based on an integral equation for the probability density function $P(E)$ giving the probability that a tunneling electron transfers the energy $E$ to the modes of the electrodynamic environment. $P(E)$ is the fundamental quantity in the theory of the DCB  and it determines the $I(U)$-curve of the tunnel junction \cite{Devoret_1990,Girvin_1990,Grabert_1991,Ingold_1992}. 

\begin{figure}[h]
\centering
{a) \hfill}

\includegraphics[width=0.4\textwidth]{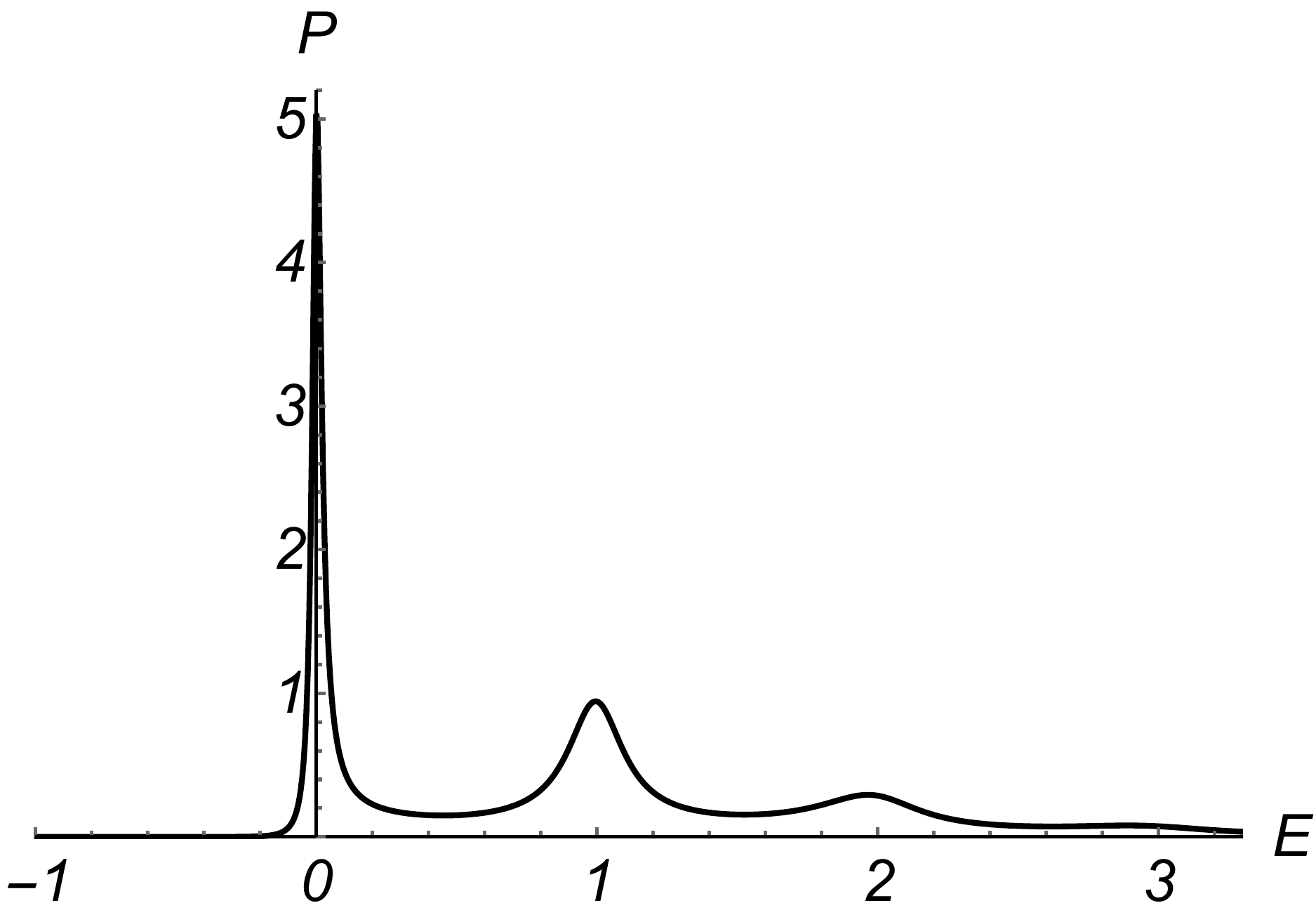}
\vspace{10pt}

{ b)\hfill}

\includegraphics[width=0.4\textwidth]{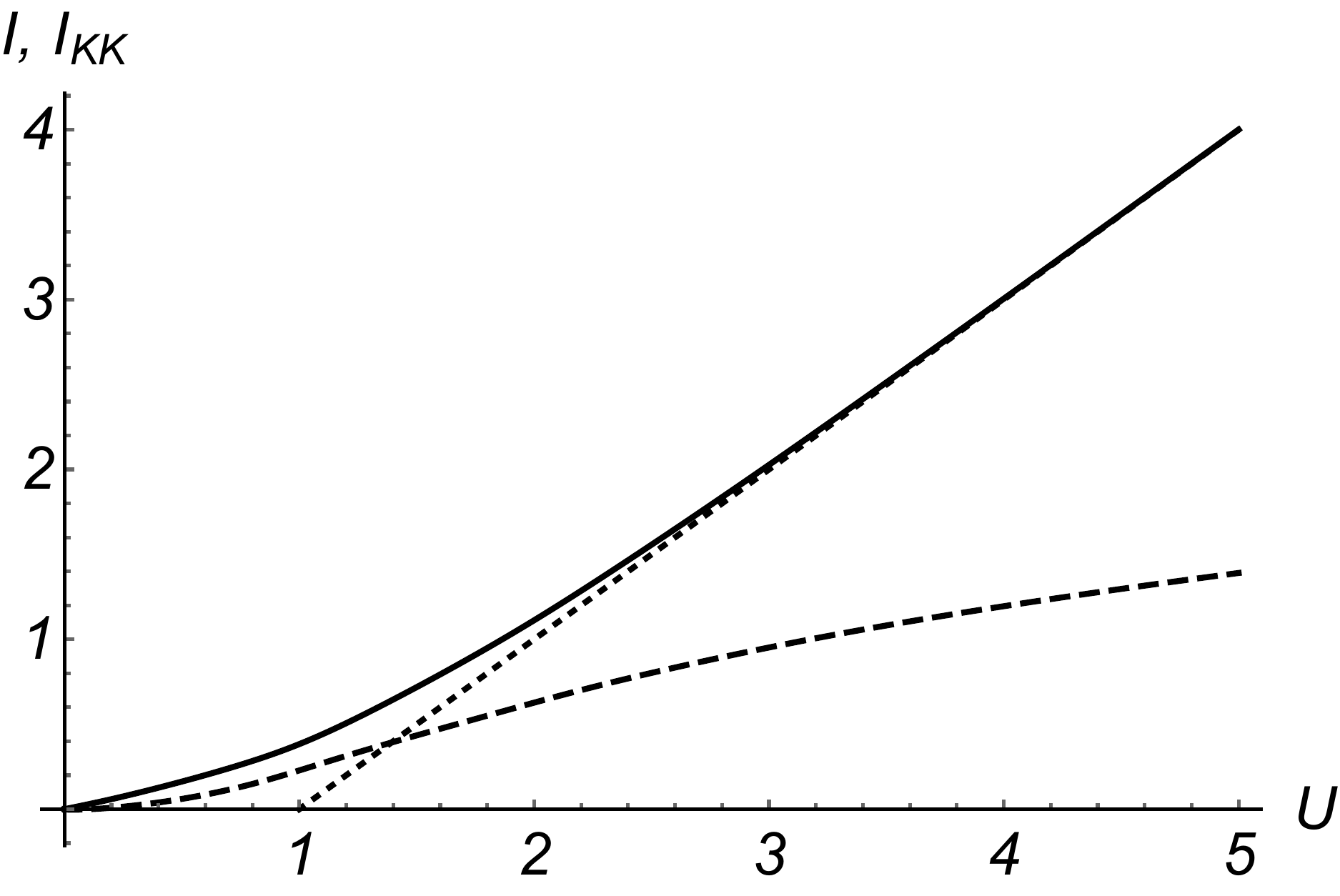}
\caption{\label{fig2} In this figure we show data for a tunnel junction driven through an $LC$-circuit with quality factor $Q_f=5$ at low temperatures ($k_BT=0.05E_c$). a) The probability density function $P(E)$. The energy is taken in units of $\hbar\omega_0$ and $P(E)$ is shown in units of $1/\hbar\omega_0$. b) The current voltage characteristic $I(U)$ (full line), its asymptote for large $U$ (dotted line) and its Kramers-Kronig transform $I_{KK}(U)$ (dashed line). The voltage $U$ is taken in units of $\hbar\omega_0/e$ and the current is shown in units of $g_T e\omega_0 $.} 
\end{figure}

We choose a temperature $T$ with $k_BT=0.05 E_c$ where $E_c=e^2/2C$ is the charging energy. DCB effects are only observable at temperatures where $k_BT$ is well below the charging energy. Furthermore, we assume a resonance frequency $\omega_0$ of the $LC$-resonator with $\hbar\omega_0=E_c$. For these parameters the method of Ref.~\cite{Ingold_1991} yields the probability density function $P(E)$ shown in Fig.~\ref{fig2}a. Apart from a peak near zero energy corresponding to almost elastic electron tunneling, $P(E)$ displays peaks at multiples of the resonance frequency which correspond to the transfer of one or more quanta of $\hbar\omega_0$ to the environment. These peaks are broadened due to the finite quality factor \cite{Ingold_1991}.

The current voltage characteristic $I(U)$ is obtained from $P(E)$ by means of a simple integration \cite{Devoret_1990,Girvin_1990,Grabert_1991,Ingold_1992}. Fig.~\ref{fig2}b depicts $I(U)$ and its Kramers-Kronig transform $I_{KK}(U)$. The asymptotic behavior of $I(U)$ for large voltages is also indicated. These currents are proportional to the tunneling conductance $G_T$ of the junction and we have introduced the dimensionless tunneling conductance
\begin{equation}
g_T=\frac{\hbar}{e^2}G_T
\end{equation}
to scale out the dependence on $G_T$ from the data shown in Fig.~\ref{fig2}b.

\begin{figure}[h]
\centering
{a) \hfill}

\includegraphics[width=0.4\textwidth]{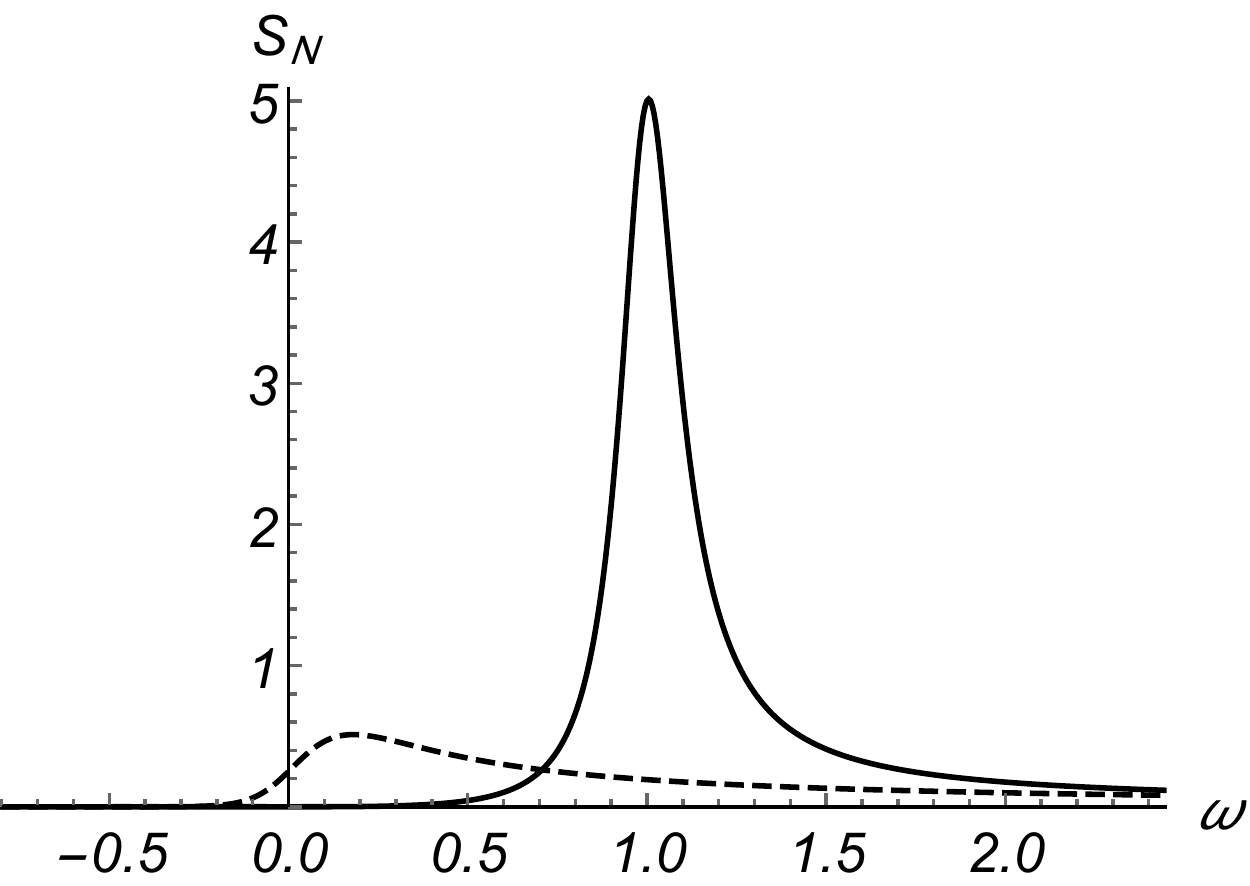}
\vspace{10pt}

{ b)\hfill}

\includegraphics[width=0.4\textwidth]{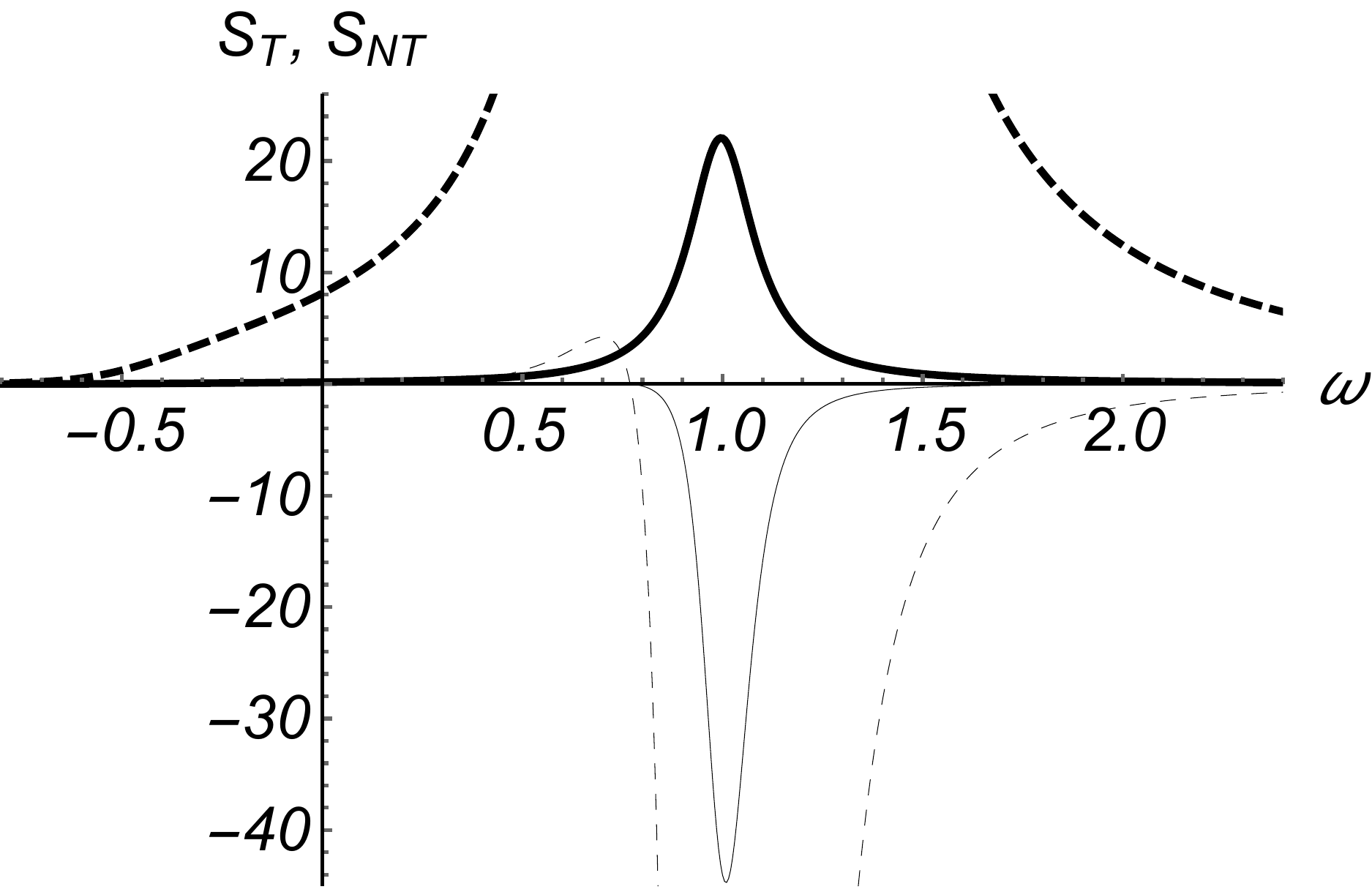}
\caption{\label{fig3} Results for the same set of parameters as in Fig.~\ref{fig2}. a) The Johnson-Nyquist part $S_N(\omega)$ of the current noise spectrum. The frequency is taken in units of $\omega_0$ and $S_N$ is shown in units of $g e^2\omega_0$. The dotted line depicts the Johnson-Nyquist noise generated by the environmental impedance. b) The shot noise part $S_T(\omega)$ (thick line) and the cross-correlation part $S_{NT}(\omega)$ (thin line) of the current noise spectrum for an applied voltage with $eU=0.5\,E_c$. The frequency is taken in units of $\omega_0$ while $S_T$ and $S_{NT}$ are shown in units of $g_T e^2\omega_0 $. The dashed lines display the spectral functions scaled up by a factor of 50.} 
\end{figure}

With the current voltage characteristic at hand, we can use Eqs.~(\ref{SNLC}), (\ref{STLC}) and (\ref{SNTLC}) to determine the spectral function of current fluctuations. The Johnson-Nyquist part $S_N(\omega)$ of the current noise spectrum is independent of the applied voltage and proportional to the lead conductance $1/R$. To scale out the dependence on $R$, we introduce the dimensionless conductance
\begin{equation}
g=\frac{\hbar}{e^2}\frac{1}{R}
\end{equation}
The spectral function $S_N(\omega)$ in units of $g e^2\omega_0$ is displayed in Fig.~\ref{fig3}a. Also shown is the Johnson-Nyquist noise of the environmental impedance (\ref{ZLC}) to illustrate the effect of the transmission factor (\ref{HC}).  The shot noise contribution $S_T(\omega)$ and the cross-correlation noise $S_{NT}(\omega)$ are shown in Fig.~\ref{fig3}b in units of $g_T e^2\omega_0$ for an applied voltage $U=E_c/2e$. The noise is strongly enhanced near the resonance frequency of the $LC$-circuit.

Since the Johnson-Nyquist noise dominates in the weak tunneling limit, it is advantageous to study the excess noise specifying the difference between the nonequilibrium current noise and its equilibrium level
\begin{equation}
S_{ex}(\omega,U)=S(\omega,U)-S(\omega,0)
\end{equation}
Here we have made the voltage dependence of the current noise spectrum explicit.  The Johnson-Nyquist part of the noise $S_N(\omega,U)$  does not contribute to the excess noise since it is independent of $U$.
\begin{figure}[h]
\centering
{a) \hfill}

\includegraphics[width=0.4\textwidth]{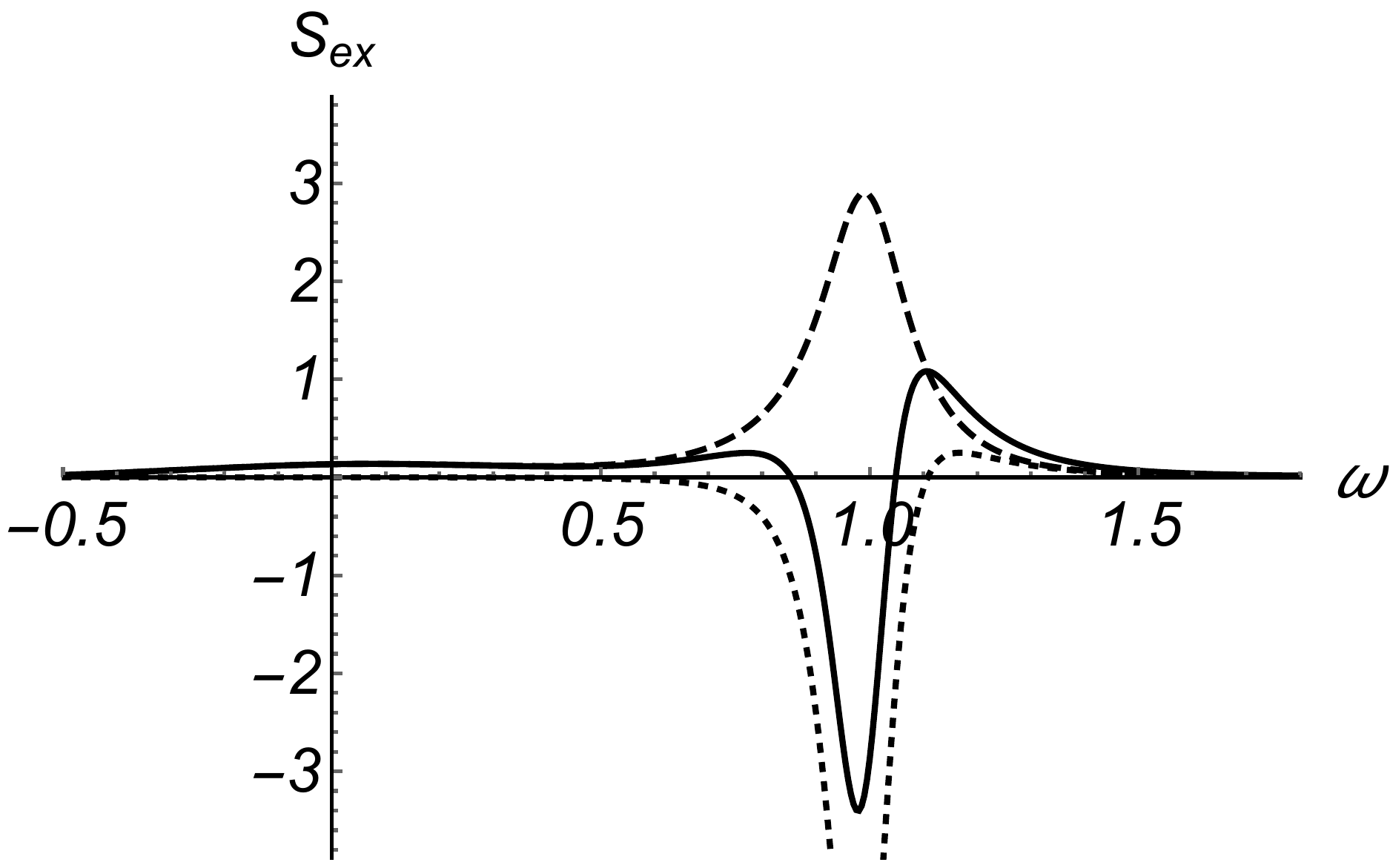}
\vspace{10pt}

{ b)\hfill}

\includegraphics[width=0.4\textwidth]{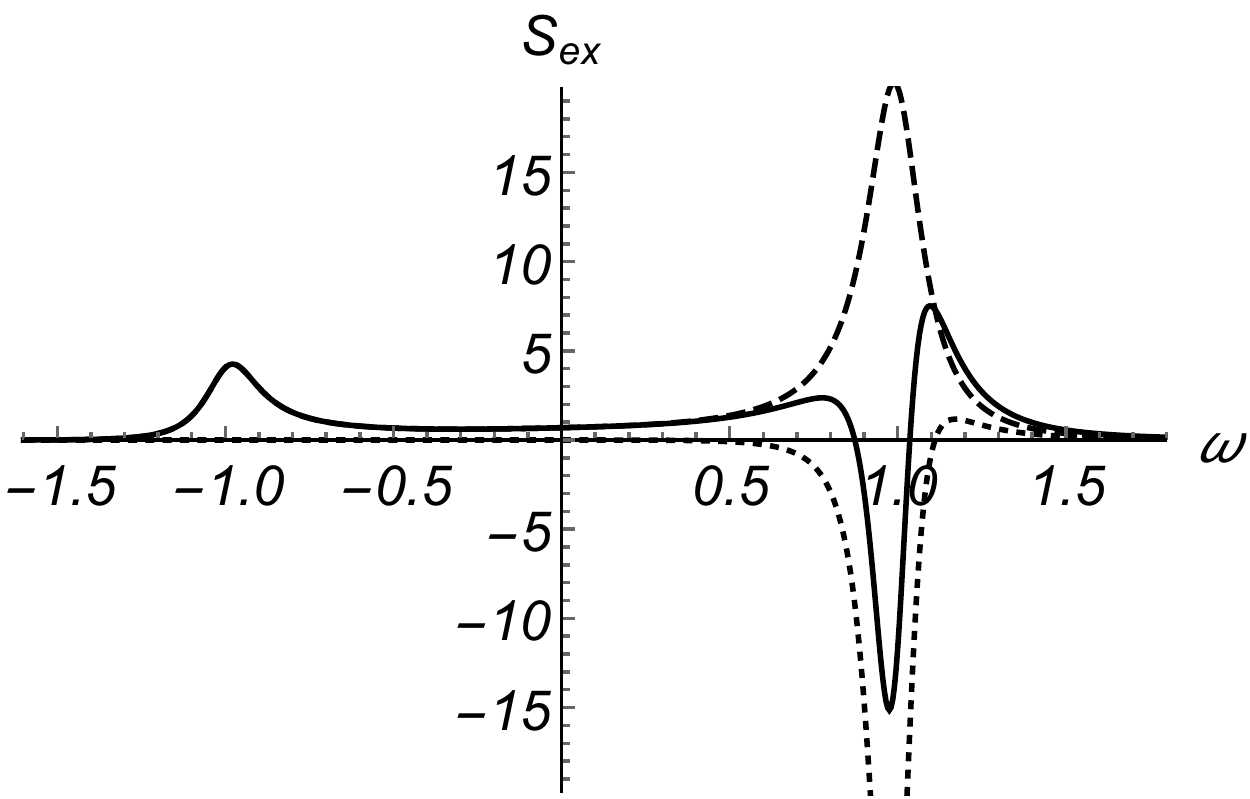}
\caption{\label{fig4} The excess noise $S_{ex}(\omega,U)$ for the same set of parameters as in Fig.~\ref{fig2} for an applied voltage of a) $U=0.5\,E_c/e$ and b) $U=1.5\,E_c/e$. The frequency is again taken in units of $\omega_0$ and $S_{ex}$ is displayed in units of $g_T e^2\omega_0$. The excess noise contributions of the shot noise (dashed curve) and of the cross-correlation noise (dotted curve) are also shown.} 
\end{figure}
In Fig.~\ref{fig4} we show results for the excess noise for two values of the applied voltage $U$. The excess noise is mostly concentrated near the resonance frequency and it grows strongly with increasing applied voltage. Furthermore, when the applied voltage $U$ exceeds $\hbar\omega_0/e$ as in Fig.~\ref{fig4}b, the shot noise part of the noise leads to a peak of the excess noise also for frequencies around $-\omega_0$. In this case the voltage source supplies energy quanta of the order of $\hbar\omega_0$ to excite the $LC$-resonator.

\section{Conclusions}\label{sec:five}
We have presented a simple derivation of the current noise spectrum of a voltage biased tunnel junction based on phenomenological considerations and circuit theory. The spectral function $S(\omega)$ of current fluctuations in the experimentally accessible outer circuit was obtained as the sum of three terms: $S_N(\omega)$ originating from the Johnson-Nyquist noise caused by the environmental impedance, $S_T(\omega)$ due to the shot noise generated at the tunnel junction and a third contribution, $S_{NT}(\omega)$, arising form a cross-correlation between fluctuations of the tunneling current and voltage fluctuations across the environmental impedance.
The Johnson-Nyquist noise $S_N(\omega)$ does not contribute to the excess noise measured relative to the equilibrium noise at vanishing applied voltage. 

We have given concrete results for a tunnel junction driven through an $LC$ resonator. The two parts of the excess noise arising from $S_T(\omega)$ and $S_{NT}(\omega)$, respectively, are found to be equally important and they are strongly enhanced near the resonance frequency of the $LC$-resonator.

In previous work \cite{Frey_2016} we have studied the spectral function $S(\omega,U,V,\Omega)$ of current fluctuations of a tunnel junction driven by an applied voltage 
\begin{equation}
V_{ext}(t)= U +V\cos(\Omega t)
\end{equation}
consisting of a dc voltage $U$ and a sinusoidal voltage of amplitude $V$ and frequency $\Omega$. The spectral function $S(\omega,U,V,\Omega)$ was shown to be determined by the spectral function $S(\omega,U)$ of the device measured under dc bias $U$ via a photo-assisted tunneling relation of the Tien-Gordon type
\begin{equation}\label{TG}
S(\omega,U,V,\Omega)=\sum_{k=-\infty}^{\infty}J_k(a)^2\, S(\omega,U+k\hbar\Omega/e)
\end{equation}
Here $J_k$ is the Bessel function of the first kind and
\begin{equation}
a=\frac{\Xi(\Omega)V}{\hbar\Omega}
\end{equation}
The factor $\Xi(\Omega)$ arises as the modulus (\ref{polar}) of the transmission factor (\ref{HZ}) at the driving frequency $\Omega$. Since the relation (\ref{TG}) determines the spectral function $S(\omega,U,V,\Omega)$ of an ac driven tunnel junction as weighted and translated copies of the spectral function $S(\omega,U)$ discussed in the previous sections, the results given there can easily be extended to ac driven systems. Yet, the relation (\ref{TG}) itself has so far only been established by means of a full-fletched calculation  \cite{Frey_2016} based on the microscopic Hamiltonian model of the DCB theory.

\section*{Appendix: Admittance of the tunnel junction}
In the presence of an external dc voltage the average tunneling current is given by
\begin{equation}
\langle I_T\rangle = I(U)
\end{equation}
When the voltage across the junction is altered by an additional ac voltage  $V_J\cos(\omega t)$ of frequency $\omega$,  the average tunneling current can be expressed in terms of the current $I(U)$ in a dc biased system by a photo-assisted tunneling relation\cite{Parlavecchio_2015}
\begin{eqnarray}\label{Idcac}\nonumber
&&\langle I_T(t) \rangle = \frac{1}{2}\sum_{k,l=-\infty}^{\infty} J_{k}(a)\\ 
&&\quad\times\Big\{ \left[J_{k+l}(a) +J_{k-l}(a)\right]
I(U+k\hbar\omega/e)\\ \nonumber
&&\quad\ +i \left[J_{k+l}(a) -J_{k-l}(a) \right]
I_{KK}(U+k\hbar\omega/e)\Big\}e^{-il\omega t}
\end{eqnarray}
Here $J_k(z)$ is the Bessel function of the first kind, $I_{KK}(U)$ is the Kramers-Kronig transform of $I(U)$ and
\begin{equation}\label{defa}
a=\frac{eV_J}{\hbar\omega} .
\end{equation}
To determine the response of the tunneling current to a small ac voltage, we expand the Bessel functions up to terms of linear order in $a$. One has\cite{Abramowitz}
\begin{equation}
J_0(a)=1+\mathcal{O}(a^2)
\end{equation}
\begin{equation}
J_1(a)=-J_{-1}(a)=\frac{a}{2} +\mathcal{O}(a^2)
\end{equation}
\begin{equation}
J_n(a)= \mathcal{O}(a^2) \quad\hbox{for}\ n \ne 0,\pm 1\, .
\end{equation}
Accordingly, we obtain from Eq.~(\ref{Idcac})
\begin{eqnarray}
&&\langle I_T(t)\rangle = I(U) \\ \nonumber
&&\qquad+\frac{a}{2}\big[I(U+\hbar\omega/e)-I(U-\hbar\omega/e)\big]\cos(\omega t)\\ \nonumber
&&\qquad+\frac{a}{2}\big[2I_{KK}(U)-I_{KK}(U+\hbar\omega/e)\\ \nonumber
&& \qquad\qquad -I_{KK}(U-\hbar\omega/e)\big]\sin(\omega t)\ +\mathcal{O}(a^2)\, .
\end{eqnarray}
This result may be rewritten to yield the linear response $\langle \delta I_T(t)\rangle = \langle I_T(t)\rangle - I(U)$ of the average tunneling current to a small ac voltage $\delta V_J(t)$
\begin{eqnarray}\nonumber
\langle \delta I_T(t)\rangle &=&\frac{e}{2\hbar\omega}\Big\{\big[I(U+\hbar\omega/e)-I(U-\hbar\omega/e)\big]\delta V_J(t)\\ 
&&\quad-\big[2I_{KK}(U)-I_{KK}(U+\hbar\omega/e) \\ \nonumber
&& \qquad-I_{KK}(U-\hbar\omega/e)\big]\frac{1}{\omega}\delta\dot V_J(t)\Big\}
\end{eqnarray}
where we have made use of Eq.~(\ref{defa}).
In Fourier space the linear response of the tunneling current takes the form
\begin{equation}
\langle \delta I_T(\omega)\rangle = Y_J(U,\omega) \delta V_J(\omega)
\end{equation}
with the junction admittance (\ref{YJ}).


\section*{Acknowledgments}
One of the authors (HG) wishes to thank the members of the Quantronics Group, CEA-Saclay, France, for inspiring discussions, frequent hospitality, and fruitful cooperation within the last thirty years.


\end{document}